\documentclass[onecolumn,showpacs,amsmath,amssymb,dvipdfm]{revtex4}

\usepackage{epsfig} 
\usepackage{amsmath}
\usepackage{amsbsy,color}

\newcommand{\beq}{\begin{equation}}
\newcommand{\eeq}{\end{equation}}
\newcommand{\beqa}{\begin{eqnarray}}
\newcommand{\eeqa}{\end{eqnarray}}

\newcommand{\psiavg}{\langle \psi \rangle}
\newcommand{\gtavg}{\langle \gamma_T \rangle}
\newcommand{\gtaavg}{\langle \langle \gamma_T \rangle \rangle}

\newcommand{\kavg}{\langle \kappa \rangle}

\newcommand{\simgt}{\lower.5ex\hbox{$\; \buildrel > \over \sim \;$}}
\newcommand{\simlt}{\lower.5ex\hbox{$\; \buildrel < \over \sim \;$}}

\def\pa{{\partial}}

\def\ha{\frac{1}{2}}
\def\imag{\dot \imath}

\def \sn2{\left(S/N\right)^2}

\begin{document} 

\title{Halo-Galaxy Lensing: A Full Sky Approach}
\author{Roland de Putter$^{1}$,
Masahiro Takada$^{2}$
\vspace{0.1cm}}

\affiliation{
$^1$Berkeley Lab \& University of California, Berkeley, CA 94720, USA\\
$^2$Institute for the Physics and Mathematics of the Universe (IPMU), 
The University of Tokyo, Chiba 277-8582, Japan
} 
\date{\today}

\begin{abstract} 
The halo-galaxy lensing correlation function or the average tangential
shear profile over sampled halos is a very powerful means of measuring
the halo masses, the mass profile, and the halo-mass correlation
function of very large separations in the linear regime. We reformulate
the halo-galaxy lensing correlation in harmonic space. We find that,
counter-intuitively, errors in the conventionally used flat-sky
approximation remain at 
a 
\% level even at very small angles.
 The errors increase at larger angles and
for lensing halos at lower redshifts: the effect is at 
a
 few \% level at
the baryonic acoustic oscillation scales for lensing halos of $z\sim
0.2$, and comparable with the effect of primordial non-Gaussianity with
$f_{\rm NL}\sim 10$ at large separations. Our results allow to
readily estimate/correct for the full-sky effect on a high-precision
measurement of the average shear profile available from upcoming
wide-area lensing surveys.
\end{abstract} 

\maketitle

\section{Introduction \label{sec:intro}}

Dark matter halos hosting galaxies and galaxy clusters emerge as a result
of gravitational amplification of tiny primordial fluctuations in the
cold dark matter (CDM) dominated structure formation scenario. The weak
lensing distortion effect on the shapes of background galaxies due to
the halos is a powerful means of measuring the 
mass 
and the mass
profile
independent of the dynamical state and the uncertain relation
between baryonic and dark matter distributions (e.g.~see
\cite{BartelmannSchneider:01} for a thorough review). Weak lensing is now
recognized as one of the most promising methods for constraining
cosmology including the nature of dark energy
(e.g.~\cite{TakadaBridle:07}),
which is the primary motivation for both on-going and planned high precision weak lensing surveys,
such as
the CFHT Legacy
Survey \footnote{http://www.cfht.hawaii.edu/Science/CFHLS/}, the Hyper
Suprime-Cam Weak Lensing Survey \cite{Miyazakietal:06}, the Dark Energy
Survey (DES) \footnote{http://www.darkenergysurvey.org}, and ultimately the
Large Synoptic Survey Telescope (LSST) \footnote{http://www.lsst.org},
Euclid \cite{Euclid:10}, and the Joint Dark Energy Mission (JDEM) \footnote{http://jdem.gsfc.nasa.gov/}.

There is a promising weak lensing technique, the so-called stacked
lensing or halo-galaxy lensing correlation, which has been increasingly
recognized as a statistically robust method since the first measurement
a decade ago (\cite{Fischeretal:00,McKayetal:01}; also see
\cite{GuzikSeljak:01,Hirataetal:04,Mandelbaumetal:05,Mandelbaumetal:06,Sheldonetal:09,Okabeetal:10}
for the recent measurements).  The halo-galaxy lensing correlation can
be measured first by measuring the tangential ellipticity/shear
component of background galaxies with respect to the halo center in each
halo region (galaxy or galaxy cluster selected beforehand) and then by
averaging the tangential shear profiles over all the sampled halos that
are tagged by a similar halo richness indicator and are in the same
redshift range.  Although lensing information for individual halos is
lost, the halo-galaxy lensing correlation has several notable advantages. First, the
average lensing profile becomes insensitive to substructures and
asphericity of the individual halos and also to the projection effect,
i.e.~the contamination arising from uncorrelated large-scale structure
along the line-of-sight. This is because these ``contaminating signals''
are averaged out via the stacking, under the assumption that the
Universe is {\em statistically} homogeneous and isotropic. The stacked
shear profile thus allows to extract the halo mass enclosed within
a three-dimensional sphere of a given radius,
which is an important quantity when using the
halo mass function for constraining cosmology.
Second, since the signal probed is linear in an estimator of lensing
shear or galaxy ellipticity, the halo-galaxy lensing is more robust to
systematic errors: the systematic errors, if relatively well behaved,
are to some extend averaged out, in contrast to the cosmic shear
measurements, where the measurement rests on the two-point correlation
based methods and therefore the systematic errors may accumulate. For
example, if a relatively clean selection of background galaxies is
available, which is usually the case for cluster lensing
\cite{Okabeetal:10}, the stacked lensing is not contaminated by the
intrinsic alignment of galaxy images, which is one of the potentially
dominant systematic errors in the cosmic shear measurement.

Thus the halo-galaxy lensing allows a precise measurement of the mean
halo mass and the mean mass profile, which is useful in calibrating the
halo mass-observable relations \cite{Rozoetal:10}.  The stacked shear
profile can boost the signal-to-noise ratios at very small angular
scales as well as very large scales, while only small solid angles or
too small signals are available for individual halo lensing.  The
stacked lensing signals on small scales arise from the mass distribution
gravitationally bound within each halo (the so-called one-halo term),
while the large-scale signals arise from the surrounding mass
distribution, such as filamentary structures, correlated with the
sampled halos (the two-halo term) (see
\cite{GuzikSeljak:01,Johnstonetal:07} for the halo model based
modeling). In particular, \cite{Jeongetal09} recently studied that, for
planned wide-area lensing surveys, the halo-mass correlation function at
large angles can be in principle used to measure features of the
baryonic acoustic oscillations and to probe the effect of primordial
non-Gaussianity \cite{Dalaletal08}, which appear at projected radii
greater than $100$ Mpc corresponding to 10 degree scales for lensing halos
at $z\sim 0.2$.

In this paper we reformulate the halo-galaxy lensing correlation
function in harmonic space, i.e.~by employing the full-sky approach,
since all the previous observational and theoretical works of
halo-galaxy lensing have been done based on the flat-sky approximation
(see \cite{Stebb96,Hu:00} for the study of the lensing power spectrum based
on the full-sky approach). We study how the full-sky correction is
relevant on angular scales probed by future wide-area surveys, and is
compared with the BAO features and with the effect of primordial
non-Gaussianity.  We indeed find a non-trivial scale-dependent
correction to the flat-sky approximation, and that counter-intuitively,
the correction is not confined to large angles.

The structure of this paper is as follows. In Sec.~\ref{sec:formulation}
we reformulate the average tangential shear profile based on the
full-sky approach.  In Sec.~\ref{sec: comp} we compare the shear profiles
computed from the full-sky and flat-sky approaches. 
Sec.~\ref{sec:conc} is devoted to brief conclusions.  Readers who want to
skip the derivations could go straight to our main analytic results
Eqs.~(\ref{eq: result 1}) and (\ref{eq: result 2.2}) and then to
Sec.~\ref{sec: comp} where we show our numerical results.  Unless
explicitly stated we assume cosmological parameters to match a flat,
$\Lambda$-dominated CDM model that is consistent with the recent WMAP
results \cite{Komatsuetal10}.

\section{Formulation}
\label{sec:formulation}
\subsection{Weak Lensing Shear on the Sky}
\label{sec: tangsh}

The bending of light by foreground mass, gravitational lensing, is
described 
by a mapping of the position of a point on the sky with coordinates $(\tilde{x}^1, \tilde{x}^2)$ (in the absence of lensing) to
a different position with coordinates $(x^1, x^2)$
(e.g. see \cite{BartelmannSchneider:01} for a thorough review).
 We
call $x^i$ 
the lens- or image-plane
coordinates and $\tilde{x}^i$ 
the source-plane coordinates. 
The deflection or displacement vector on the sky
 is given by the gradient
of a {\it lensing potential} $\psi$:
\beq
\tilde{x}^i = x^i - \psi^{;i},
\eeq
where
$\psi$ is a line of sight projection of the 
gravitational potential
and
``$;$'' denotes a contravariant derivative 
with respect to the image coordinates. For a finite-size source such as a
galaxy, lensing causes its image to be distorted.
The distortion 
is 
described by
differences in deflection angles
between different parts of  a source galaxy image. 
The small displacement
  vector connecting the two points in the
observed (or lens-plane) coordinates, 
$\delta x^i$, is mapped from the vector in the source plane, $\delta
\tilde{x}^i$, as 
\beq
\label{eq: gen trans}
\delta \tilde{x}^i = \left(\delta^i_{\,\, j} 
- 
\psi_{\,\,\, ;j}^{;i} \right) \delta x^j.
\eeq
Although the above expressions are covariant and valid in any coordinate system,
to define the shear components $\gamma_1$ and $\gamma_2$, it is convenient to work
in a locally orthonormal system,
i.e.~$g_{ij} = \delta_{ij}$, where $\delta_{ij}$ is the Kronecker delta function.
In such a coordinate system, where one does not have to distinguish between lower and upper
indices, we define the shear by
\beq
\delta {\bf \tilde{x}} =
\begin{pmatrix}
\ 1 - \kappa - \gamma_1 & -\gamma_2 \ \\
\ -\gamma_2 & 1 - \kappa + \gamma_1 \ 
\end{pmatrix}\, \delta {\bf x},
\eeq
with
\beqa
\label{eq: kg1g2}
\kappa 
&\equiv& 
\ha \nabla^2 \psi, \nonumber \\
\gamma_1 
&\equiv& 
\ha \left(\psi_{;1 1} - \psi_{;2 2}\right), \\
\gamma_2 
&\equiv& 
\psi_{;1 2}. \nonumber
\eeqa
The convergence is a particularly informative quantity because
it is a direct line of sight integral of the matter density fluctuation field
$\delta_m$
(see appendix \ref{sec: app1}).

The tangential shear around a halo is in the flat sky approach defined
in a polar coordinate system with the halo at the origin. On the spherical sky, this needs to be generalized
to a spherical coordinate system. 
Without loss of generality
the halo center can be taken at the north pole due to the statistical
isotropy of the Universe.
In spherical coordinates, the metric is given by
\beq
d s^2 = d \theta^2 + \sin^2\theta\, d \phi^2.
\eeq
At each point on the sphere, this coordinate system defines a local basis $\left({\bf e}_{\theta}, {\bf e}_{\phi} \right)$
which is orthogonal, but not orthonormal. Orthonormality can be realized by locally making the transformation to the coordinates
$(\theta, \varphi)$, where $\varphi$ is defined by $d \varphi = \sin \theta\, d \phi$.
We then define the tangential and transverse shears by in the $(\theta, \varphi)$ system setting
\beqa
\label{eq: gT def}
\gamma_T \equiv - \gamma_1, \quad \gamma_\times \equiv \gamma_2 \quad \quad {\rm (in }\,\,(\theta, \varphi)\,\,{\rm system)}.
\eeqa

It will be useful to have expressions for $\kappa$, $\gamma_T$ and $\gamma_\times$ in terms of derivatives of the lensing potential
with respect to $\theta$ and $\phi$. To derive these, we will first
construct $\psi_{;i j}$ in the $(\theta, \phi)$
system and then use the fact that $\psi_{;i j}$ is a rank two tensor to transform to the
$(\theta, \varphi)$ system.
We thus start by calculating the covariant derivatives of $\psi$ with respect to $\theta$ and $\phi$.
The first covariant derivative is a derivative of a scalar and therefore simply equals the partial derivative:
$\psi_{;i} = \psi_{,i}$.
The second derivative is the derivative of the covariant vector $\psi_{;i}$ and is given by
\beq
\psi_{;i j} = \psi_{, i j} - \Gamma^k_{i j} \psi_{,k},
\eeq
where the Christoffel symbols are defined by
\beq
\Gamma^a_{bc} = \ha g^{a d} \left( g_{dc ,b} + g_{bd ,c} - g_{bc ,d}\right).
\eeq
The non-vanishing Christoffel symbols are 
\beqa
\Gamma^{\theta}_{\phi \phi} = -\sin\theta \cos\theta \nonumber \\
\Gamma^\phi_{\theta \phi} = \Gamma^{\phi}_{\phi \theta} = \frac{\cos\theta}{\sin\theta}
\eeqa
and all others Christoffel symbols are zero.
This gives
\beq
\psi_{;i j} =
\begin{pmatrix}
\  \pa_\theta^2 \psi & \sin\theta \pa_\theta \left(\frac{1}{\sin \theta} \pa_\phi \psi\right) \ \\
\ \sin\theta \pa_\theta \left(\frac{1}{\sin \theta} \pa_\phi \psi\right) & \pa_\phi^2 \psi + \sin\theta \cos\theta \pa_\theta \psi \ 
\end{pmatrix}, \quad (i, j = \theta, \phi).
\eeq
We now transform to the orthonormal system, 
where the basis is transformed as ${\bf e}_\phi \cdot 
{\bf e}_\phi  \neq 1 \rightarrow {\bf e}_\varphi \cdot{\bf e}_\varphi = 1 $, 
giving
\beq
\label{eq: 2nd der}
\psi_{;i j} =
\begin{pmatrix}
\  \pa_\theta^2 \psi & \pa_\theta \left(\frac{1}{\sin \theta} \pa_\phi \psi\right) \ \\
\ \pa_\theta \left(\frac{1}{\sin \theta} \pa_\phi \psi\right) & \frac{1}{\sin^2\theta} \pa_\phi^2 \psi + \frac{\cos\theta}{\sin\theta} \pa_\theta \psi \ 
\end{pmatrix}, \quad \quad (i, j = \theta, \varphi).
\eeq
Using Eqs.~(\ref{eq: kg1g2}) and (\ref{eq: gT def}), we can now easily identify
\beqa
\label{eq: kg vs psi}
\kappa &=& \ha \left( \psi_{;\theta \theta} + \psi_{;\varphi \varphi}\right)
= \ha \left(\pa_\theta^2 \psi + \frac{1}{\sin^2\theta} \pa_{\phi}^2 \psi + \frac{\cos\theta}{\sin\theta} \pa_\theta \psi\right) \nonumber \\
\gamma_T &=& - \ha \left( \psi_{;\theta\theta} 
- \psi_{;\varphi\varphi} \right)
=  -\ha \left( \pa_{\theta}^2 \psi - \frac{1}{\sin^2\theta} \pa_{\phi}^2 \psi - \frac{\cos\theta}{\sin \theta} \pa_\theta \psi \right) \\
\gamma_{\times} &=& \psi_{;\theta \varphi} = \pa_\theta \left( \frac{1}{\sin\theta} \pa_\phi \psi \right). \nonumber
\eeqa
Eqs.~(\ref{eq: kg vs psi}) will be the starting point of the derivations of the relation between tangential shear and
projected mass and the relation between tangential shear and the angular power spectrum of the halo-mass cross correlation.

\subsection{Tangential Shear}
\label{sec: spher Kaiser}

A quantity of great cosmological interest is the angle-averaged tangential shear $\gtavg(\theta)$,
where in general the angle average of a quantity $\alpha$ is defined as
\beq
\label{eq: angle avg}
\langle \alpha \rangle (\theta) \equiv \frac{1}{2 \pi} \int_0^{2 \pi} d \phi \, \alpha(\theta, \phi).
\eeq
It follows from rotational symmetry that after stacking, there is no loss of information if we restrict ourselves
to the angle-averaged shear and convergence $\gtavg (\theta)$, $\langle \gamma_\times \rangle (\theta)$
and $\langle \kappa \rangle (\theta)$.

A useful property of $\langle \gamma_\times \rangle (\theta)$ and $\gtavg(\theta)$ is that the former
has to equal zero and the latter can conveniently be expressed in terms of the projected matter density.
In the flat sky approximation \cite{KaiserSquires:93,SquiKais96,Miralda-Escude:96},
\beq
\label{eq: kaiser}
\gtavg(\theta) = \bar{\kappa}( < \theta) - \kavg(\theta),
\eeq
where
\beq
\label{eq: avg2}
\bar{\kappa}(< \theta) = \frac{2}{\theta^2} \int_0^\theta d\theta' \, \theta'\, \kavg(\theta')
\eeq
is the mean convergence within a circular disk with angular radius $\theta$.
The convergence in turn is a direct line of sight integral of the
matter overdensity (see appendix \ref{sec: app1}).

We will now derive the full-sky equivalent of Eq.~(\ref{eq: kaiser}).
Using Eqs.~(\ref{eq: kg vs psi}) and Eq.~(\ref{eq: angle avg}), we find
\beqa
\label{eq: gtk18}
\langle \gamma_T \rangle(\theta) = - \ha \left( \pa_{\theta}^2 \langle \psi \rangle - \frac{\cos\theta}{\sin\theta} \pa_\theta \psiavg \right) \nonumber \\
\kavg (\theta)= \ha \left( \pa_{\theta}^2 \psiavg + \frac{\cos \theta}{\sin\theta} \pa_\theta \psiavg \right).
\eeqa
To relate the shear to the convergence, we need to define the 
mean convergence averaged within circular aperture of radius
$\theta$ (the full-sky version of Eq.~(\ref{eq: avg2})),
\beqa
\label{eq: kavg}
\bar{\kappa}(< \theta) &\equiv& \frac{1}{2 \pi (1 - \cos\theta)}
\int_0^\theta 
\sin\theta' \,d\theta' 
\int_0^{2 \pi} 
d\phi \, ~ \kappa(\theta, \phi) \nonumber \\
&=& \frac{1}{1 - \cos\theta} \int_0^{\theta} \sin\theta' \, d\theta' \kavg (\theta') \nonumber \\
&=& \frac{\sin\theta}{2\left(1 - \cos\theta\right)} \pa_\theta \psiavg.
\eeqa
Eqs.~(\ref{eq: gtk18}) and (\ref{eq: kavg}) above now allow us to write
\beq
\label{eq: result 1}
\gtavg(\theta) = 2 \frac{\cos\theta}{1 + \cos\theta}\, \bar{\kappa}(< \theta) - \kavg(\theta).
\eeq
This is the main result of this subsection. Note that in the small-angle limit, $\theta \ll 1$,
the flat-sky result (\ref{eq: kaiser}) is recovered.

\subsection{Average Tangential Shear Profile: 
Halo-Mass Correlation Function}
\label{sec: main}

An even more useful observable is the so-called ``{\em stacked
tangential shear}'' or ``{\em halo-galaxy correlation function}'', 
which is obtained by computing the tangential shear
profile of background galaxy images relative to each halo~\footnote{
Here we mean by ``relative to'' that we calculate the tangential shear in a coordinate system
that has the halo at the north pole.
},
observationally represented by a galaxy or cluster, and then taking the
average over all the sampled halos 
\cite{Fischeretal:00,Sheldonetal:01,McKayetal:01}. The stacked shear
profile can be estimated as
\beq
\langle \gtavg \rangle (\theta) \equiv \frac{1}{N_h} \sum_{\rm halos} \gtavg(\theta),
\eeq
where $N_h$ is the total number of halos averaged over.
The ensemble average of this quantity gives the halo-shear
correlation function or the halo-mass correlation function: 
\beq
\langle \gtavg \rangle (\theta) = 
\langle \delta_h(\vec{\theta}')\, \gamma_T(\vec{\theta}'+\vec{\theta})
\rangle
 = \langle \delta_h(0) \, \gtavg(\theta) \rangle,
\eeq
where 
$\delta_h(\vec{\theta})$ is the angular density fluctuation field of
halos on the sky, and 
$\gamma_T$ is defined relative to the position $(\theta', \phi')$. The independence of $\theta'$ and $\phi'$
follows from rotational symmetry, which
comes from the statistical isotropy of the universe. 

By comparing the observed stacked 
shear to the value predicted by theory, one can learn about large scale structure
and the background expansion history
(e.g. see \cite{GuzikSeljak:01} for such 
a study).
To calculate the theoretical $\gtaavg$, one first uses the halo-mass cross power
spectrum $P^{hm}(k, z)$ and the background cosmology to construct the angular convergence-halo spectrum $C_l^{\kappa h}$
(see appendix \ref{sec: app1}).
Then, in the flat sky approximation, this angular spectrum is used to calculate $\gtaavg$ as a function of $\theta$
through
\beq
\label{eq: flat}
\gtaavg(\theta) = \int \frac{l d l}{2 \pi}\, 
C_l^{h\kappa} 
J_2(l \theta),
\eeq
where 
$C_l^{h\kappa}$ is the angular power spectrum of cross-correlation
between halos and
the convergence field due to mass distribution surrounding the halos
for a given source galaxy population, 
and 
$J_n$ is the
$n$-th order
 Bessel function of the first kind. However, this relation is modified
when the curvature of the sky is taken into account. 
Below we will derive the exact relation.

We start by expanding the halo
 density fluctuation field
and 
the lensing potential field
in spherical harmonics:
\begin{eqnarray}
\delta_h(\theta, \phi) &=& \sum_{l, m} a_{lm}^h \, Y_{l m}(\theta, \phi),
\nonumber\\
\psi(\theta, \phi)& =& \sum_{l, m} a_{lm}^\psi \, Y_{l m}(\theta, \phi).
\label{eq: psi harm}
\end{eqnarray}
The spherical harmonics are
\beq
\label{eq: spher harm}
Y_{l m}(\theta, \phi) = \sqrt{\frac{2 l + 1}{4 \pi}\frac{(l - m)!}{(l + m)!}} \, P_l^m(\cos\theta) \, e^{\imag m \phi},
\eeq
with $P_l^m$ the associated Legendre polynomials.
The angular power spectra are in general defined by
\beq
\langle \left(a_{lm}^{X}\right)^* a_{l'm'}^Y\rangle = C_l^{XY} \delta_{ll'} \delta_{mm'}.
\eeq

The tangential shear can according to Eqs.~(\ref{eq: kg vs psi}) and (\ref{eq: psi harm}) be expanded as
\beq
\gamma_T = - \ha \sum_{l m} a_{l m}^{\psi} \left(Y_{lm ;\theta \theta}(\theta, \phi) - Y_{lm ; \varphi \varphi}(\theta, \phi) \right).
\eeq

It is now straightforward to work out the stacked tangential shear:
\beqa
\label{eq: first step}
\gtaavg(\theta) &=&  \langle \delta^h(0) \, \gamma_T(\theta, \phi)\rangle \nonumber \\
&=& - \ha \sum_{lm} Y_{lm}^*(0) \left( Y_{lm ; \theta \theta}(\theta, \phi) - Y_{lm ; \varphi \varphi}(\theta, \phi) \right) \,C_l^{\psi h} \nonumber \\
&=& - \ha \sum_{l} \sqrt{\frac{2 l + 1}{4 \pi}} \left( Y_{l 0 ; \theta \theta}(\theta, \phi) - Y_{l 0 ; \varphi \varphi}(\theta, \phi) \right) \,C_l^{\psi h}
\eeqa
(using $Y_{lm}(0) = \delta_{m, 0} \, \sqrt{(2 l + 1)/4 \pi}$).
The second (covariant) derivatives of a scalar in the $(\theta, \varphi)$ coordinate system
can be read off from Eq.~(\ref{eq: 2nd der}) or (\ref{eq: kg vs psi}),
giving for the terms in parentheses
\beqa
\label{eq: ylm0}
Y_{l 0 ;\theta \theta} (\theta, \phi) - Y_{l 0 ; \varphi \varphi}(\theta, \phi) &=&
\sqrt{\frac{2 l + 1}{4 \pi}} \, \left(\pa_{\theta}^2 - \frac{\cos \theta}{\sin \theta} \pa_\theta \right) P_l(\cos \theta) \nonumber \\
&=& \sqrt{\frac{2 l + 1}{4 \pi}} \, \left( \sqrt{1 - x^2} \frac{d}{dx}\left( \sqrt{1 - x^2} \frac{d}{dx} P_l(x)
\right) + x \frac{d}{dx} P_l(x)\right) \nonumber \\
&=& \sqrt{\frac{2 l + 1}{4 \pi}} \, (1 - x^2) \frac{d^2}{dx^2} P_l(x) \nonumber \\
&=& \sqrt{\frac{2 l + 1}{4 \pi}} P_l^2(x) , \quad x \equiv \cos\theta,
\eeqa
where $P_l$ is the Legendre polynomial and $P_l^2$ is the {\it associated} Legendre polynomial with $m=2$.
Since $\kappa = \ha \nabla^2 \psi$ and $\nabla^2 Y_{lm} = -l(l + 1) Y_{lm}$,
\beq
\label{eq: cl}
C_l^{\psi h} = - \frac{2}{l (l + 1)} \, C_l^{\kappa h}.
\eeq
Inserting 
Eq.~(\ref{eq: ylm0}) and 
Eqs.~(\ref{eq: cl}) into (\ref{eq: first step}) gives the final result
\beq
\label{eq: result 2.2}
\gtaavg(\theta) = \sum_l C_l^{\kappa g} \frac{2l + 1}{4 \pi \, l (l + 1)} P_l^2(\cos\theta).
\eeq

In the small-angle, large-$l$ limit \cite{Stebb96}, $l \gg 1, \theta \ll 1$,
\beq
P_l^m(\cos\theta) \approx (-1)^m \frac{(l + m)!}{(l - m)!} \, l^{-m}
J_m(l \theta),
\label{eq:Pl_limit}
\eeq
so that for small angles, the 
flat-sky
result (\ref{eq: flat}) is recovered.

Eq.~(\ref{eq: result 2.2}) is the main result of this paper as it gives the exact expression
for the observed stacked 
shear profile
in terms of the theoretical angular power spectrum of the mass-halo
correlation.

\section{Results: comparing full- and flat-sky approaches }
\label{sec: comp}

\begin{figure*}[!htb]
\begin{minipage}[t]{0.49\textwidth}
\centering
  \includegraphics*[width=\linewidth]{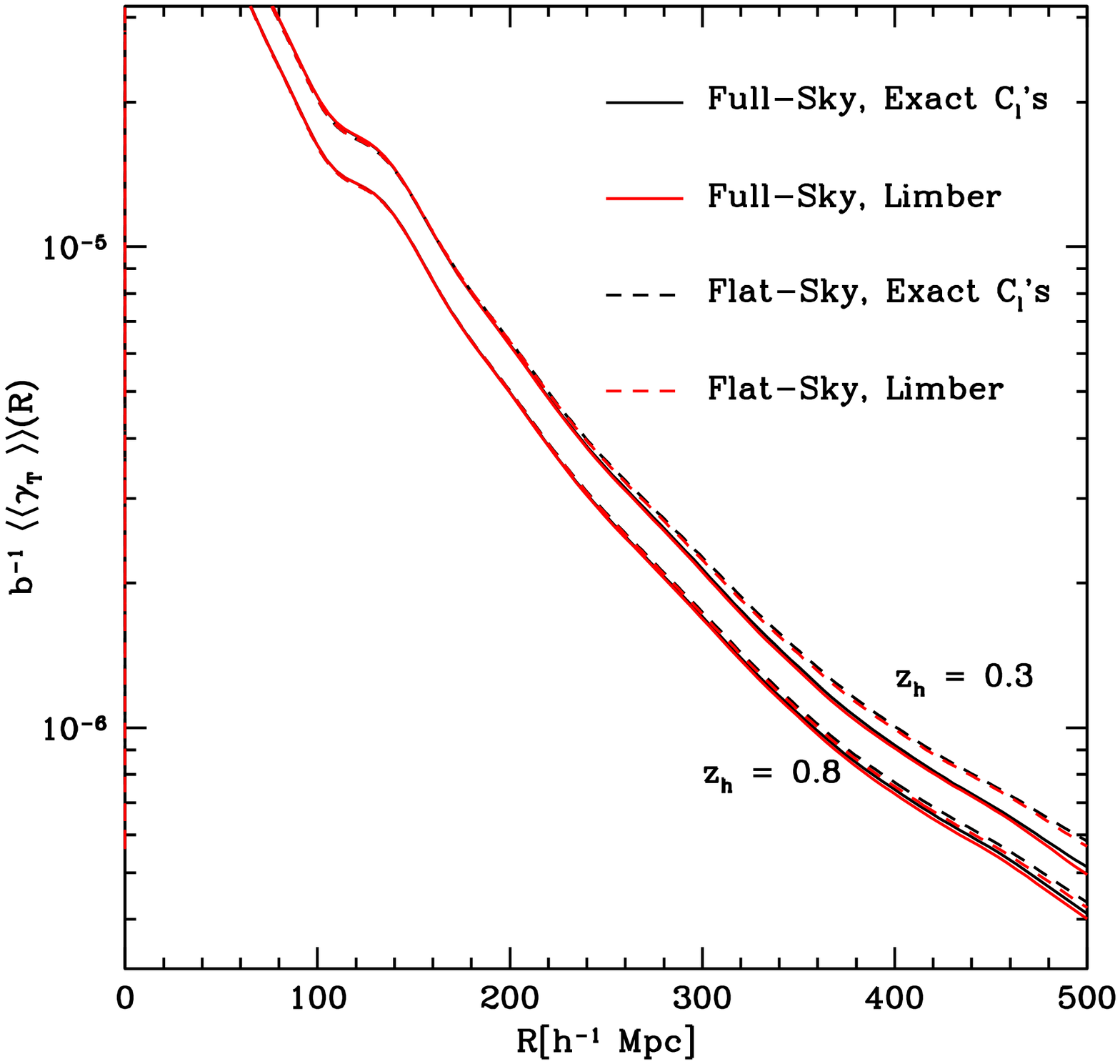}
\end{minipage} \hfill
\begin{minipage}[t]{0.49\textwidth}
\centering
  \includegraphics*[width=\linewidth]{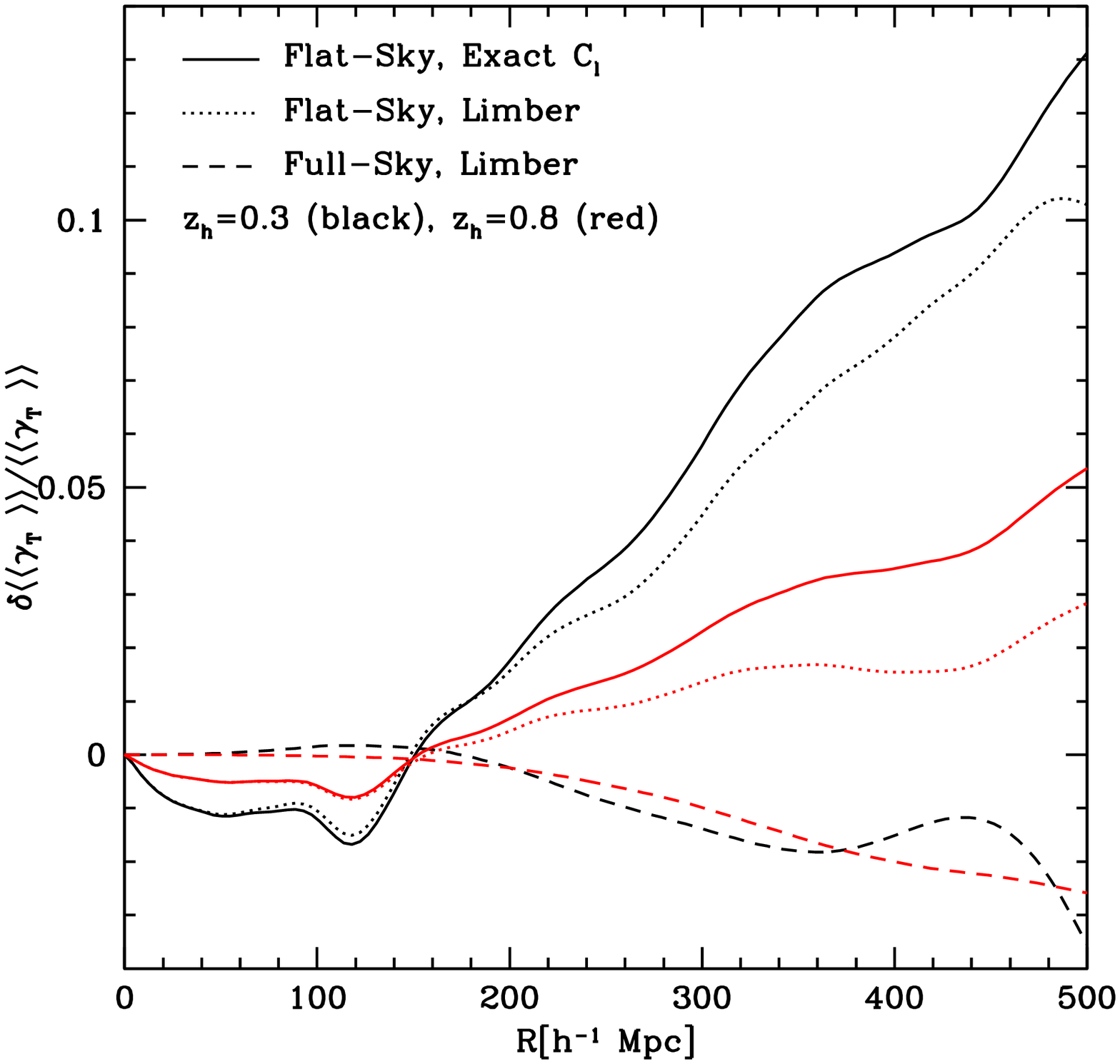}
\end{minipage} \hfill
  \caption{
{\em Left panel}: The average tangential shear profile (divided by bias)
as a function of halo centric radii for lensing halos 
at redshifts $z_h=0.3$ and $0.8$. The solid curves are the results
 computed from the full-sky approach, developed in this paper, while
 the dashed curves show the results from the flat-sky approximation. The
 curves denoted by ``Limber'' show the results using the
 Limber approximation for the calculation of the halo-mass power spectrum (see text for
 the details). 
  {\em Right panel}: Relative differences between various approximations
  and the exact full-sky result. Note that the difference is defined as
$\delta\langle\langle\gamma_T\rangle\rangle/\langle\langle\gamma_T\rangle\rangle
\equiv \langle\langle\gamma_T\rangle\rangle_{\rm flat}/
\langle\langle\gamma_T\rangle\rangle_{\rm full} -1.
$
  }
  \label{fig: g_T}
\end{figure*}

Now that we have an exact, full-sky expression for 
the stacked shear profile, 
 $\gtaavg(\theta)$ 
(see Eq.~\ref{eq: result 2.2}),
we compare it to the result 
computed from 
the conventionally used flat-sky approximation.
We do this by studying $\gtaavg$
for halos at 
two lensing
redshifts $z_h = 0.3$ 
and $ 0.8$.
For the redshift distribution of source galaxies we assumed $z_0 \approx 0.37$
for the
functional form $n(z)\propto z^2 \exp[-z/z_0]$
given by Eq.~(7) in \cite{TakJain04} such that the mean source redshift
$\langle z\rangle=1$. Then we used the galaxies sitting in the redshift
range $z=[1,1.5]$ as for source galaxies, yielding the mean redshift $\langle z\rangle\simeq
1.2$.
For the matter power spectrum needed to compute the angular power
spectrum $C_l^{h\kappa}$ we used the non-linear power spectrum, obtained by applying
the HaloFit prescription \cite{Smithetal03} to the linear power spectrum, for a
concordance $\Lambda$CDM model that is consistent with the WMAP 7-year
results \cite{Komatsuetal10}.
For the halo bias we simply assume a linear bias throughout this paper: the halo-mass
correlation strengths are different from the mass correlation strengths
by some constant factor.
Although a more precise modeling of the halo-galaxy lensing requires to
use a suite of N-body simulations, our approach is adequate enough,
because the primary purpose of this paper is to study the impact of
full-sky 
approach being compared with the conventionally used flat-sky
results.

In Fig.~\ref{fig: g_T}
we study
the stacked shear 
profile
as a function of
comoving transverse distances 
from the halo center, 
$R \equiv \theta \, \chi_h$. 
Note that the comoving radial distances to halos at $z_h=0.3$ and $0.8$
are $\chi_h=840$ and $1984$ $h^{-1}$Mpc, respectively, for a flat
$\Lambda$CDM model and therefore the
comoving transverse length 
of
$100$ $h^{-1}$Mpc corresponds to 6.82 and 2.89
degrees, respectively, for the two lensing redshifts. 
We also study
the effect of the Limber approximation \cite{Limber:54}, which 
is often used in the literature
 to calculate the angular spectrum of halo-mass correlations
$C_l^{h \kappa}$ (see Appendix~\ref{sec: app1} for the details). 

The left panel of Fig.~\ref{fig: g_T} compares the average tangential
shear profile, computed from the full-sky approach, with those computed from
various approximations, while the right panel shows the relative
differences. 
The three approximations we consider are: flat-sky shear with exact $C_l^{h \kappa}$, flat-sky shear with
Limber approximated $C_l^{h \kappa}$, 
and
full-sky shear with Limber approximated $C_l^{h \kappa}$.
The main interest of this paper is the comparison of the flat-sky approximation to the exact expression,
so we focus now on the flat-sky with exact $C_l^{h \kappa}$ case.
The first thing to notice is that the error caused by the flat-sky approximation
is larger for $z_h = 0.3$ than $z_h=0.8$, 
which can be explained by noting that a
fixed 
transverse
comoving distance 
to the line-of-sight
corresponds
to a larger angle 
for lower lensing redshifts. 
 As for the scale-dependence of the difference,
as expected, it starts at zero for $R=0$. Then, it starts to deviate
linearly from zero 
even at small distances,
with the flat-sky expression underestimating the signal. 
The perhaps counter-intuitive errors at small angles arise from the
leading order correction in the asymptotic approximation relating
the 2nd-order Legendre polynomial $P^2_l$ to the 2nd-order Bessel
function in Eq.~(\ref{eq:Pl_limit}), as explained in detail
in Appendix \ref{sec: app2}. 
The difference reaches the $1 \%$ level around the BAO scale
and then 
becomes greater at larger distances.
The flat-sky expression overestimates the signal by more than  $10 \%$
at $R \simgt 400~ h^{-1}$Mpc for $z_h=0.3$.
Thus we need to account for the full-sky effect for 
a \%-level high-precision measurement of the lensing-halo
correlation, expected from upcoming wide-area lensing surveys. 

\begin{figure*}[!htb]
\begin{minipage}[t]{0.49\textwidth}
\centering
  \includegraphics*[width=\linewidth]{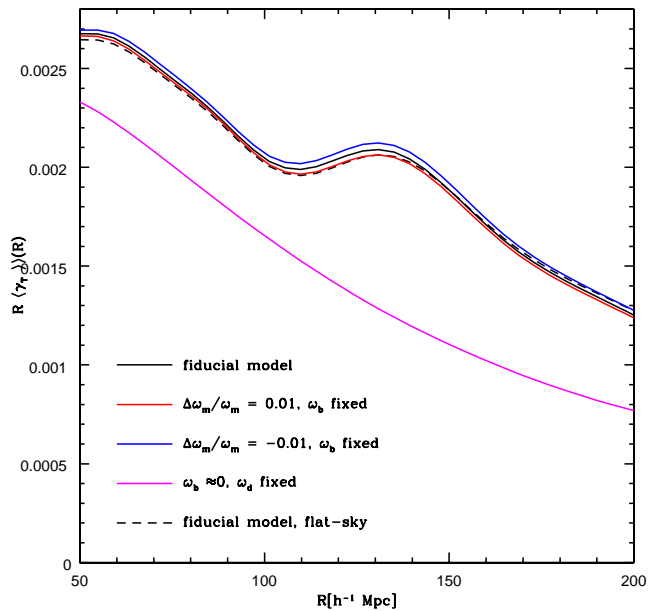}
\end{minipage} \hfill
  \caption{The BAO feature in the stacked tangential shear 
profile (divided by bias) for $z_h = 0.3$. We show the shear rescaled by a factor $R$
  to make the feature come out more clearly. The red and blue lines correspond to a variation of $\pm 1 \%$
  in the total matter density with the baryon density
kept fixed. The magenta curve corresponds to
  $\omega_b=0$, giving no acoustic feature at all. The dashed line is for the fiducial model, but using
  the flat-sky approximation.}
  \label{fig: BAO}
\end{figure*}

Recently \cite{Jeongetal09} studied that the halo-galaxy lensing profile
can be used to 
probe the BAO feature,
appearing at 
$R \approx 147$Mpc, and to explore the 
effect of primordial non-Gaussianity, which become manifest at even larger scales. Since the
error caused by the flat-sky approximation becomes significant at 
larger angles, 
it is useful
to compare this error to these signals.

In Fig.~\ref{fig: BAO},
we show the stacked tangential shear profile (divided by bias) for $z_h = 0.3$,
zooming in on the BAO feature. The figure shows that the flat-sky
error is comparable to a change in signal due to a variation of about $1 \%$ in the total matter density $\omega_m$
(i.e.~$1 \%$ of the fiducial value for $\omega_m$),
with the baryon density $\omega_b$ being fixed.

\begin{figure*}[!htb]
\begin{minipage}[t]{0.49\textwidth}
\centering
  \includegraphics*[width=\linewidth]{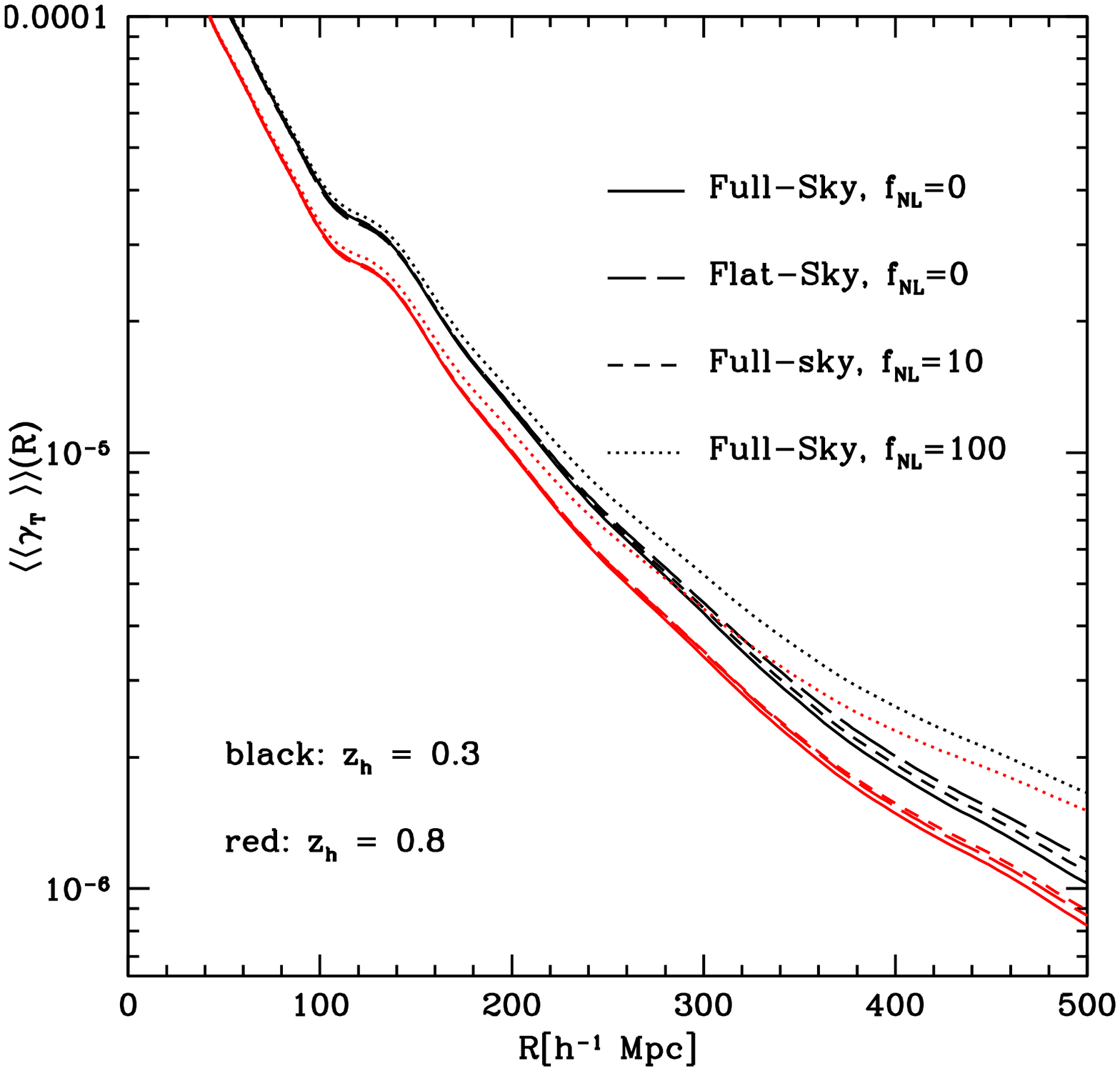}
\end{minipage} \hfill
\begin{minipage}[t]{0.49\textwidth}
\centering
  \includegraphics*[width=\linewidth]{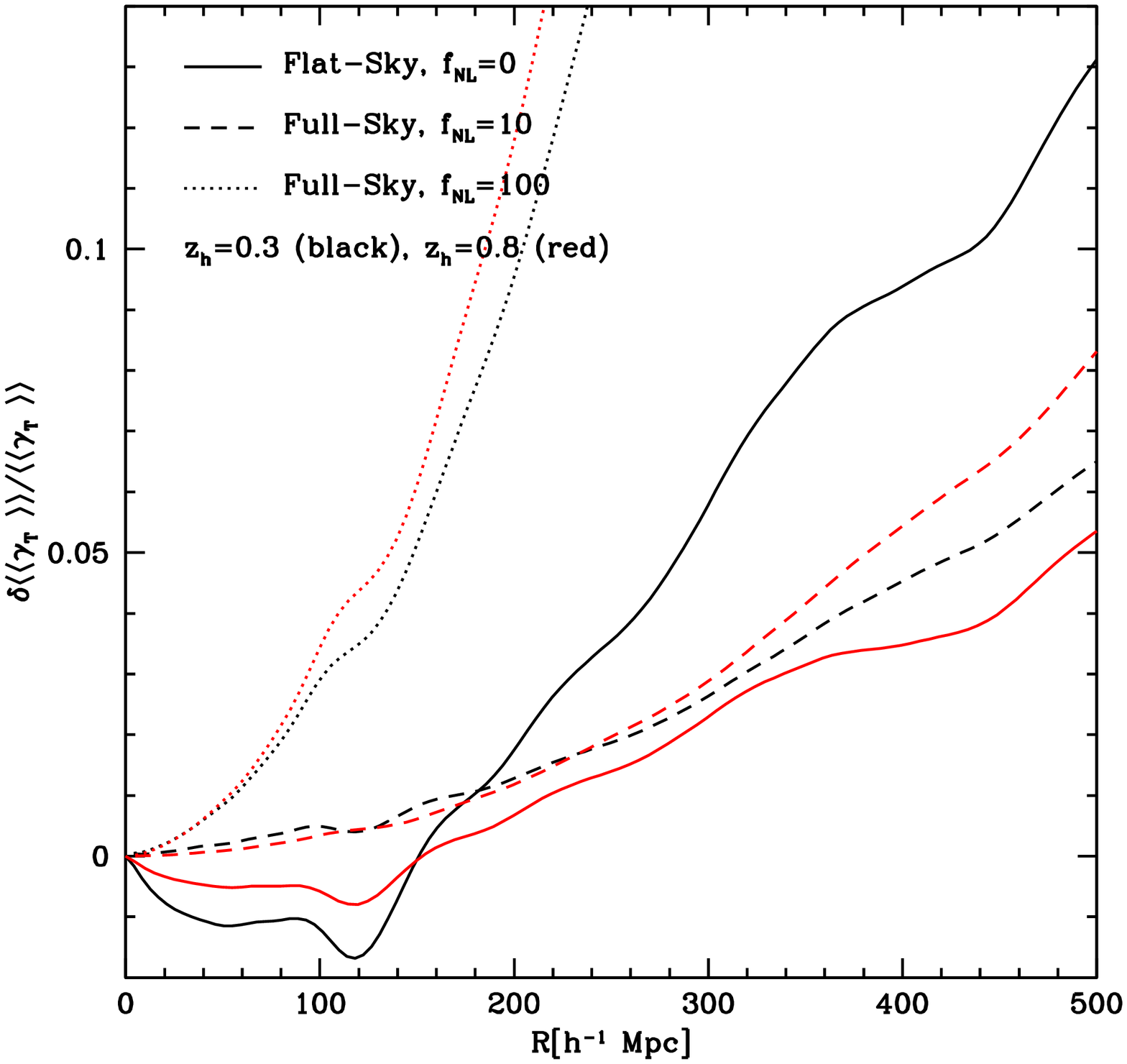}
\end{minipage} \hfill
  \caption{The stacked tangential shear profile as a function of $f_{\rm NL}$ compared
  to the flat-sky result. We assume a linear bias $b_1 = 2$. 
{\em Left panel}: The profiles themselves. 
  {\em Right panel}: 
The difference relative to the $f_{\rm NL} = 0$, exact result.
  The error due to the flat-sky approximation is comparable to the effect of
  $f_{\rm} = \mathcal{O}(10)$.}
  \label{fig: f_nl}
\end{figure*}

The local-type
primordial non-Gaussianity parametrized 
by $f_{\rm NL}$
causes a scale-dependent bias in the
halo-mass correlation 
\cite{Dalaletal08, MatVer08, Slosetal08, AfshTol08, Taruyaetal08, Desjetal09, Pille10, Grossietal09}:
the halo-mass cross power spectrum can be formally given as
\beq
\label{eq: bias}
P^{h m}(k) = \, \left[
 b_1 + \Delta b(k) \right] P^{m m}(k),
\eeq
where $P^{m m}(k)$ is the matter power spectrum.
The scale-dependent bias is given by
\beqa
\Delta b(k) 
&\equiv& f_{\rm NL} \, \left(\frac{k_{\rm NG}(a)}{k}\right)^2,
\eeqa
where
\beq
k_{\rm NG}(a) = \sqrt{\frac{3 \, (b_1 - 1) \, \Omega_{m, 0} \, \delta_c}{a \, g(a)}} \, H_0,
\eeq
which is roughly equal to the Hubble scale.
In the above, $\delta_c = 1.68$ is the critical overdensity for spherical collapse and
$b_1$ is the bias in the absence of non-Gaussianity.
The function $g(a)$ is the linear growth factor \cite{Linder:05}: 
it is well approximated by the fitting formula
\beq
g(a) = e^{\int \frac{da}{a} \, \left((\Omega_m(a))^\gamma - 1\right)},
\eeq
with $\gamma \approx 0.55$ (in GR, with a cosmological constant).

Since $C_l^{h \kappa}$ is a projection of $P^{h m}(k)$ (see 
Appendix \ref{sec: app1}),
it follows from Eq.~(\ref{eq: bias}) that $f_{\rm NL}$ affects the low $l$ part of the
angular spectrum and therefore the stacked tangential shear at large separations.
Assuming $b_1 = 2$, in Fig.~\ref{fig: f_nl} (left panel), we show the signal for $f_{\rm NL} = 0, \pm 10, \pm 100$
and compare it to the flat-sky signal for $f_{\rm NL} = 0$. In the right panel, we show the difference
of non-zero $f_{\rm NL}$ and of the flat-sky expression for $f_{\rm NL} = 0$, relative to
the exact result for $f_{\rm NL} = 0$. The error due to the flat-sky expression corresponds roughly
to a change in $f_{\rm NL}$ of order $10$.

\section{Conclusions}
\label{sec:conc}

In this paper, we 
have developed a full-sky formalism for the halo-galaxy lensing
correlation or the stacked lensing shear profile. 
We have updated the 
the flat-sky formula
relating the tangential
shear to convergence and found the exact expression for the stacked tangential
shear $\gtaavg$ in terms of the angular halo-mass cross power spectrum.
For the stacked tangential shear, we have confirmed by direct
calculation
 that errors caused by
the flat-sky approximation, which is currently used in the literature,
are very small on scales relevant for current data
($\lesssim 1 \%$ for comoving distance $R < 20 h^{-1}$Mpc). However, if
in the future
halo-galaxy
lensing is used to constrain primordial non-Gaussianity
by measuring the stacked tangential shear at separations of several hundreds
of Mpc, the error due to the flat-sky approximation becomes relevant, it being of order
$10 \%$ around $R = 400 h^{-1}$Mpc.

The size of the errors due to the flat-sky approximation, as shown in
Figs.~\ref{fig: g_T}, \ref{fig: BAO} and \ref{fig: f_nl}, can be
compared with Fig.~6 in \cite{Jeongetal09}, which shows that the
fractional errors in measuring the halo-galaxy correlation function at
these large scales for an all-sky lensing survey are at about 20-30\%
level. Therefore the errors are well within the statistical measurement
errors for existing and future lensing surveys. 
Nevertheless the exact, full-sky calculation is just as easy to do
as the flat-sky approximation, our results can give a way to correct for
or estimate the full-sky effect for future wide-area lensing surveys.

\bigskip
\noindent{\bf Acknowledgement}: We thank A.~Stebbins for useful
discussion.  This work is supported in part by World Premier
International Research Center Initiative (WPI Initiative), MEXT, Japan.
RdP has been
supported in part by the Director, Office of Science, Office of High Energy
Physics, of the U.S.\ Department of Energy under Contract
No.\ DE-AC02-05CH11231.

\appendix

\section{angular spectra and Limber approximation}
\label{sec: app1}

\begin{figure*}[!htb]
\begin{minipage}[t]{0.49\textwidth}
\centering
  \includegraphics*[width=\linewidth]{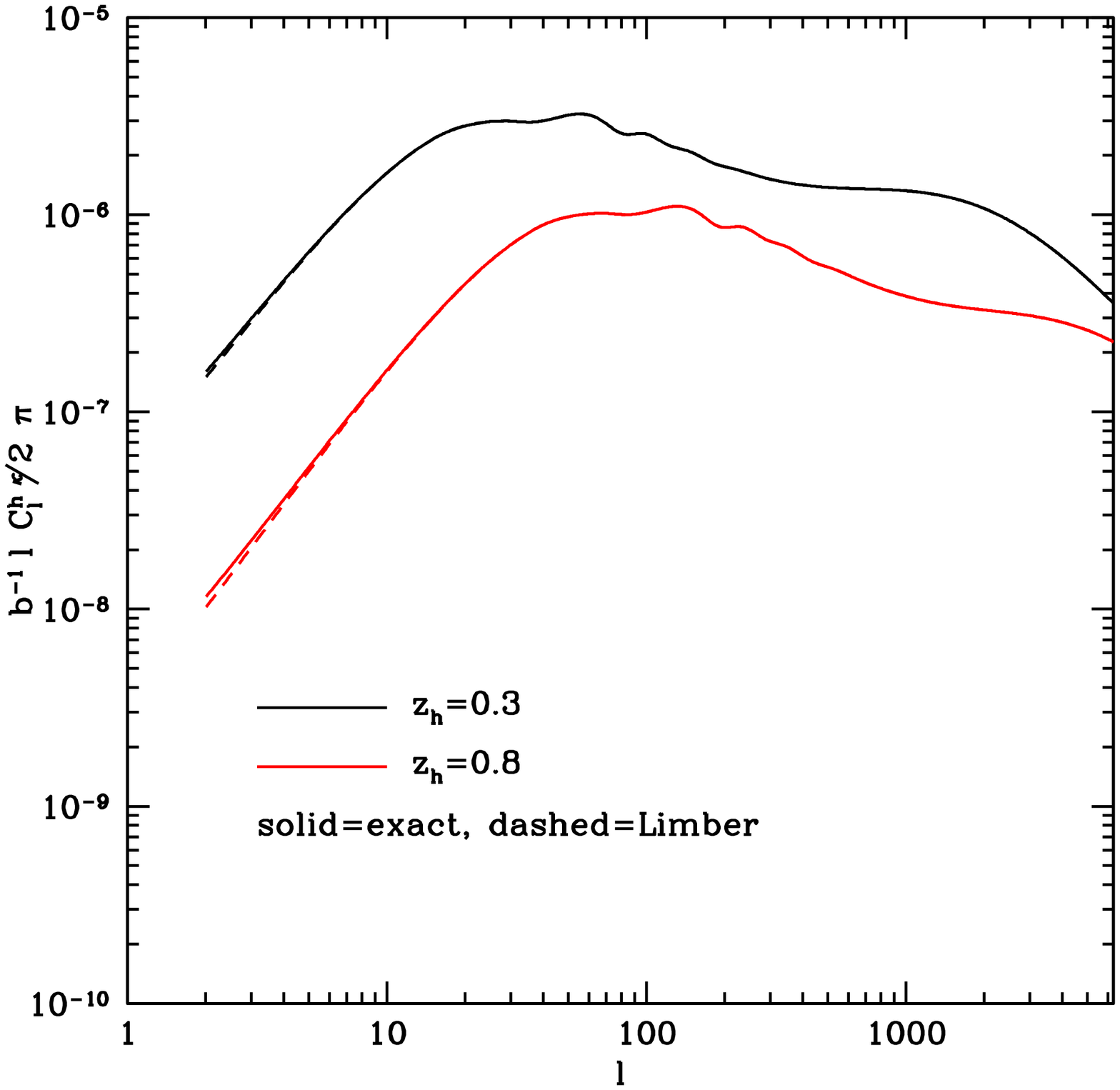}
\end{minipage} \hfill
\begin{minipage}[t]{0.49\textwidth}
\centering
  \includegraphics*[width=\linewidth]{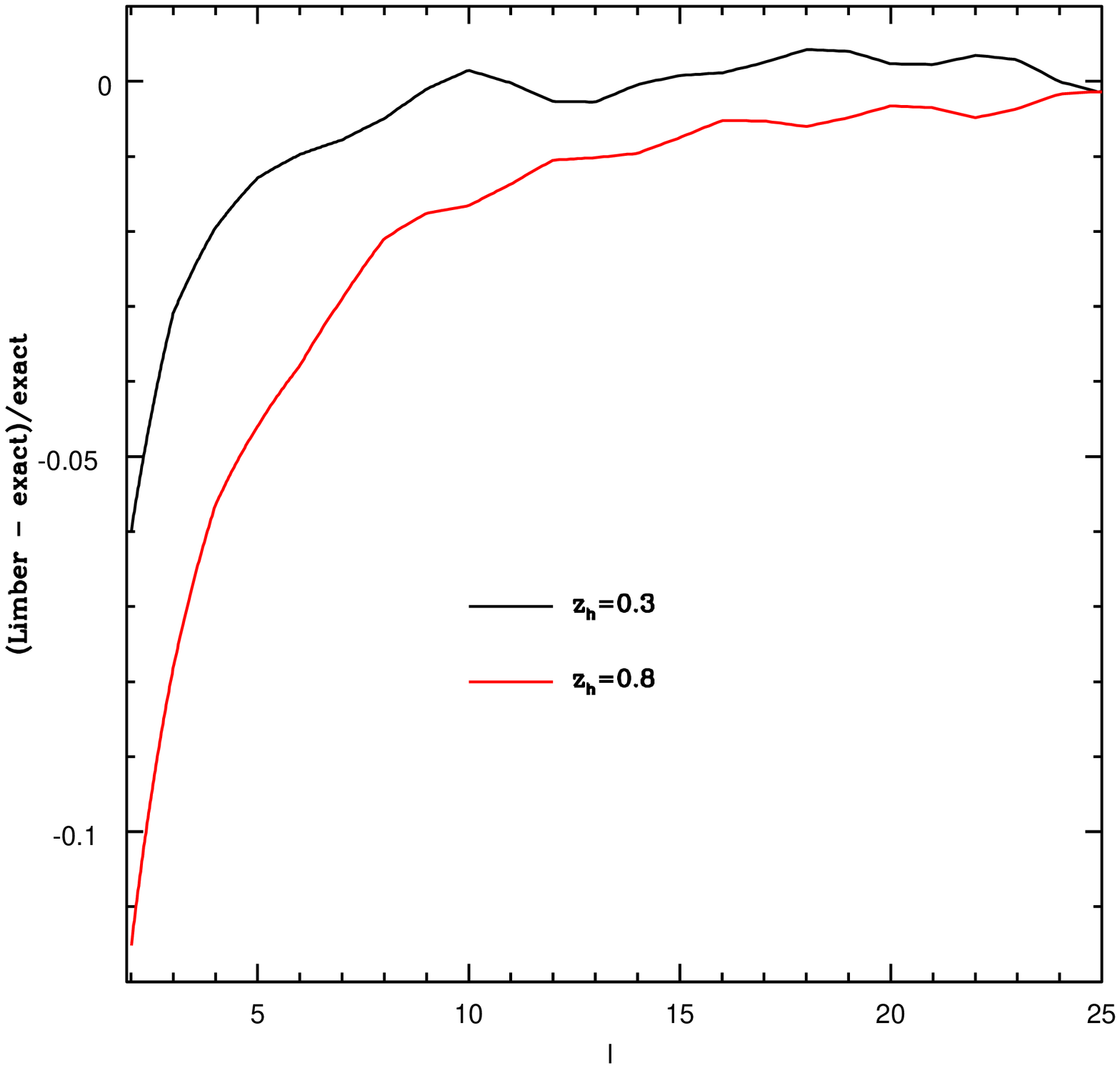}
\end{minipage} \hfill
  \caption{{\em Left panel}:
The full-sky result for the angular power spectrum of convergence and
 halo cross-correlation (divided by the linear halo bias)
is compared with the result obtained from the
 Limber approximation for lensing redshifts $z_h=0.3$ and $z_h=0.8$
 (see 
Eqs.~[\ref{eq: exact}] and [\ref{eq: Limber}]). 
  {\em Right}: The relative differences.}
  \label{fig: Cl}
\end{figure*}

To calculate $\gtaavg$, we need the angular cross spectrum of 
weak lensing convergence $\kappa$
with the 
(projected) halo density fluctuation field
$\delta_h$. These quantities can both be expressed
as line-of-sight projections of three dimensional quantities (the matter overdensity $\delta_m$ and
the halo overdensity $\delta_h$ respectively):
\beq
\kappa(\hat{n}) = \int d\chi \, W^{\kappa}(\chi) \,\delta_m(\chi \hat{n}, \tau_0 - \chi),
\eeq
and
\beq
\delta_h(\hat{n}) = \int d\chi \, W^{h}(\chi) \,\delta_h(\chi \hat{n}, \tau_0 - \chi),
\eeq
with weight functions
\beq
W^{\kappa}(\chi) = \frac{3}{2} H_0^2 \,\Omega_{m, 0} \,\chi \,(1 + z) \int_z^\infty dz_s \, \frac{\chi_s - \chi}{\chi_s} \,\left(\frac{n_s(z_s)}{n_{s, {\rm tot}}}\right),
\eeq
and
\beq
W^{h}(\chi) = H(z) \,\frac{n_h(z)}{n_{h, {\rm tot}}},
\eeq
where
$n_s(z_s)$ is the distribution of source galaxies, $\int dz_s \, n_s(z_s) = n_{s, {\rm tot}}$,
and $n_h(z)$ is the distribution of lens halos, $\int dz\, n_h(z) = n_{h, {\rm tot}}$.
Here $\chi$ is the comoving distance along the line of sight and we
assume spatial flatness.

After expanding these fields in terms of spherical harmonics,
the exact angular cross spectrum is given by
\beq
C_l^{h \kappa} = \frac{1}{2 \pi^2} \int d\chi \, W^h(\chi) \int d\chi' \, W^{\kappa}(\chi') \int d^3 \vec{k}\,
P^{h m}(k, \tau_0 - \chi, \tau_0 - \chi') \,j_l(k \chi)\, j_l(k \chi'),
\eeq
where $P^{h m}(k, \tau, \tau') \,\delta(\vec{k}' - \vec{k}) \equiv
\langle \delta_h^*(\vec{k}, \tau) \,\delta_m(\vec{k}', \tau')\rangle$,
and $j_l(x)$ denotes the $l$-th order spherical Bessel function. 

When we can obtain a cross-correlation of the lensing convergence with halos
at a single redshift
$z_h$ (and comoving distance $\chi_h$),
i.e.~$n_h(z)/n_{h, {\rm tot}} = \delta(z - z_h)$,
the above expression reduces to
\beq
\label{eq: exact}
C_l^{h \kappa} = \frac{1}{2 \pi^2} \int d^3 \vec{k} \, j_l(k \chi_h) \int d\chi \, j_l(k \chi) W^{\kappa}(\chi) P^{h m}(k, \tau_0-\chi_h, \tau_0-\chi).
\eeq
Expression (\ref{eq: exact}) is the one we use for our calculation of
the exact halo-galaxy lensing correlation.

In the Limber approximation, the expression simplifies to
\beqa
\label{eq: Limber}
C_l^{h \kappa} &=& H_0^2 \, \frac{W^{\kappa}(\chi_h)}{\chi_h} \, P^{h m}\left(\frac{l + \ha}{\chi_h}, \tau_0 - \chi_h\right) \nonumber \\
&=& \frac{3}{2} H_0^2 \,\Omega_{m, 0} \,(1 + z_h)\, \chi^{-1}_h \, P^{h m}\left(\frac{l + \ha}{\chi_h}, \tau_0 - \chi_h\right)
\eeqa

We show both the exact angular and the Limber approximated spectra (divided by bias) for $z_h=0.3$ and $z_h=0.8$
in the left panel of Fig.~\ref{fig: Cl}. In the right panel, we show the relative difference of the Limber spectra with the
exact spectra. For low $l$, the Limber approximation underestimates the spectrum by up to $12 \%$ (for $z_h=0.3$).
The difference becomes negligible at $l \sim 25$.

\section{Leading order correction for the flat-sky tangential shear profile}
\label{sec: app2}

The exact stacked tangential shear is given by
\beq
\gtaavg(\theta) = \sum_l C_l^{\kappa g} \frac{2l + 1}{4 \pi \, l (l + 1)} P_l^2(\cos\theta)
\eeq
and the
flat-sky approximation gives
\beq
\gtaavg(\theta) = \int \frac{l d l}{2 \pi}\, C_l^{\kappa g} J_2(l \theta).
\eeq
The special functions involved are related by
\beq
P_l^2(\cos\theta) = \frac{(l + 2)!}{(l - 2)!} \, l^{-2} J_2(l \theta) \left(1 + \epsilon(l, \theta)\right),
\eeq
where $\epsilon$ is small and goes to zero as $\theta \to 0$, $l \to \infty$.
For small $\theta$, the exact shear can thus be written
\beqa
\gtaavg(\theta) &=& \sum_l C_l^{\kappa g} \frac{2l + 1}{4 \pi l (l + 1)} P_l^2(\cos\theta) \\
&=& \left\{
\sum \frac{1}{2 \pi} \,\frac{l+ \ha}{l (l + 1)}\,
\frac{(l+2)!}{(l-2)!}\,l^{-2}\,J_2(l \theta) \,C_l \right\} \left(
1 + \mathcal{O}(\epsilon)\right) \\
&=& \left\{
\sum \frac{1}{2 \pi} \, l\, J_2(l \theta)\, C_l \,\left(1 +
\frac{3}{2} l^{-1} + \mathcal{O}(l^{-2}) \right) \right\}
\left(1 + \mathcal{O}(\epsilon)\right) \\
& \approx & \left\{
\int \frac{l dl}{2 \pi} \, J_2(l \theta) \, C_l \, \left(1 + \frac{3}{2} l^{-1} + \mathcal{O}(l^{-2}) \right) \right\} \left(1 + \mathcal{O}(\epsilon)\right).
\eeqa
Since the main contributions to the sum/integral come from $l \sim \theta^{-1}$, the correction to the flat-sky expression
of order $l^{-1}$ inside the integral translates
to a shear correction proportional to $\theta$. This effect alone would imply that the flat-sky approximation underestimates the shear at very low $\theta$.
However, we also need to consider the $\mathcal{O}(\epsilon)$ term.
We checked numerically that for fixed $l$,
$\epsilon \propto \theta^2$ at least as long as $l \, \theta \ll 1$. However, the main contribution
comes from $l \, \theta \sim 1$. We checked that for fixed $l \, \theta = x$ of order unity, $\epsilon = \epsilon(x/\theta, \theta) \propto \theta$ and $\epsilon < 0$
so that this
leads to an additional $\mathcal{O}(\theta)$ correction to the flat-sky shear. This effect on its own would imply that the flat-sky expression
overestimates the true shear. We find numerically that the first effect wins (see
Fig.~\ref{fig: g_T}) so that for low $\theta$, the relative difference between the flat-sky result and the exact result
is negative and evolves linearly in $\theta$.

\bibliography{refs}

\end{document}